\documentclass[aps,preprint,showpacs,eqsecnum,amssymb]{revtex4-1}
\usepackage{graphicx,psfrag}
\newcommand{\media}[1]{\left\langle  #1 \right\rangle}
\begin{document}
\date{\today}
%\flushright{\Large DRAFT}
\title{
Bound particle coupled to two thermostats}
\author{Hans C. Fogedby}
\email{fogedby@phys.au.dk}

\affiliation{Department of Physics and Astronomy, University of
Aarhus\\Ny Munkegade, 8000, Aarhus C, Denmark\\}

\affiliation{Niels Bohr Institute\\
Blegdamsvej 17, 2100, Copenhagen {\O}, Denmark}

\author{Alberto Imparato}
\email{imparato@phys.au.dk} \affiliation{Department of Physics and
Astronomy, University of Aarhus\\Ny Munkegade, 8000, Aarhus C,
Denmark}

%\author{C. Mejia-Monasterio}
%\email{} \affiliation{}

\begin{abstract}
We consider a harmonically bound Brownian particle coupled to two
distinct heat reservoirs at different temperatures. We show that
the presence of a harmonic trap does not change the large
deviation function from the case of a free Brownian particle
discussed by Derrida and Brunet and Visco. Likewise, the
Gallavotti-Cohen fluctuation theorem related to the entropy
production at the heat sources remains in force. We support the
analytical results with numerical simulations.
\end{abstract}
.

\maketitle
%%%%%%%%%%%%%%%%%%%%%%%%%%%%%%%%%%%%%%%%%%%%%%%%%%%%%%%%%
%%%%%%%%%%%%%%%%%%%%%%%%%%%%%%%%%%%%%%%%%%%%%%%%%%%%%%%%%
%%%%%%%%%%%%%%%%%%%%%%%%%%%%%%%%%%%%%%%%%%%%%%%%%%%%%%%%%
\section{\label{intro} Introduction}
%%%%%%%%%%%%%%%%%%%%%%%%%%%%%%%%%%%%%%%%%%%%%%%%%%%%%%%%%
%%%%%%%%%%%%%%%%%%%%%%%%%%%%%%%%%%%%%%%%%%%%%%%%%%%%%%%%%
%%%%%%%%%%%%%%%%%%%%%%%%%%%%%%%%%%%%%%%%%%%%%%%%%%%%%%%%%
There is a strong current interest in the thermodynamics and
statistical mechanics of small fluctuating non-equilibrium systems.
The current focus stems from the recent possibility of direct
manipulation of nano-systems and bio-molecules. These techniques
permit direct experimental access to the probability distribution
for the work and indirectly the heat distribution
\cite{Trepagnier04,Collin05,Seifert06a,Seifert06b,Wang02,
Imparato07,Ciliberto06,Ciliberto07, Ciliberto08}. These methods have
also opened the way to the experimental verification of the recent
fluctuation theorems, which relate the probability of observing
entropy-generated trajectories, with that of observing
entropy-consuming trajectories
\cite{Jarzynski97,Kurchan98,Gallavotti96,Crooks99,Crooks00,
Seifert05a,Seifert05b,Evans93,Evans94,Gallavotti95,
Lebowitz99,Gaspard04,Imparato06,
vanZon03,vanZon04,vanZon03a,vanZon04a,Seifert05c,Rondoni07,Chetrite08}.

We shall here focus on the Gallavotti-Cohen fluctuation theorem
\cite{Gallavotti95} which establishes a simple symmetry for the
large deviation function $\mu$ for systems arbitrarily far from
thermal equilibrium. Close to equilibrium linear response theory
applies and the fluctuation theorem becomes equivalent to the
usual fluctuation-dissipation theorem relating response and
fluctuations \cite{Reichl98,Kurchan98}.

A simple example of  non-equilibrium system has been introduced
recently by  Derrida and Brunet \cite{Derrida05}. In this model a
particle or rod is coupled to two heat reservoirs at different
temperatures. We also note that Van den Broeck and co-workers
\cite{vdenbroeck04,vdenbroeck08} have shown that an asymmetric
object coupled to two heat reservoirs is able to rectify the random
thermal fluctuations and thus exhibits a net motion along a
preferred direction. It is therefore of interest to know whether the
global behavior of these fluctuations, e.g., their fundamental
symmetries, are left unaltered in the case one includes a potential
or a particular interaction in such simple models. Furthermore, one
is interested in knowing what type of interaction or lattice
potential may increase, for example, the efficiency of a Brownian
motor. When dealing with systems coupled to different heat baths,
e.g., a chain of coupled oscillators, one of the main trends is to
understand which essential properties of the microscopic dynamics
lead to a diffusive limit for the energy \cite{Lepri03}. Finally, it
is also of importance to understand how heat conduction is affected
when one deals with very small systems.

More precisely, for a system driven into a steady non-equilibrium
state by the coupling to for example two distinct heat reservoirs or
thermostats at temperatures $T_1$ and $T_2$, a heat flux $dQ/dt$ is
generated in order to balance the energy. The heat flux is
fluctuating and typically its mean value $d\langle Q\rangle/dt$ is
proportional to the temperature difference. Focusing on the
integrated heat flux, i.e., the heat $Q(t)=\int_0^td\tau
(dQ(\tau)/d\tau)$ over a time span $t$, this quantity also
fluctuates and typically grows linearly in time at large times. For
the probability distribution we obtain the asymptotic long time
behavior
\begin{eqnarray}
P(Q,t)\propto e^{tF(Q/t)}, \label{dis}
\end{eqnarray}
defining the large deviation function $F(q)$. The Gallavotti-Cohen
fluctuation theorem then establishes the symmetry
\begin{eqnarray}
F(q)-F(-q)=q[1/T_1-1/T_2]. \label{ft1}
\end{eqnarray}
Likewise, for the characteristic function
\begin{eqnarray}
\langle e^{\lambda Q(t)}\rangle\propto e^{t\mu(\lambda)},
\label{char}
\end{eqnarray}
the fluctuation theorem states the symmetry relation
\begin{eqnarray}
\mu(\lambda)=\mu(-\lambda+1/T_1-1/T_2). \label{ft2}
\end{eqnarray}

The fluctuation theorem has been demonstrated under quite general
and somewhat abstract conditions \cite{Gallavotti95}. It is
therefore of importance to discuss the theorem in the context of
specific models where the large deviation function $\mu(\lambda)$
can be derived explicitly.

The large deviation function $\mu(\lambda)$ can be determined
explicitly for the simple non-equilibrium model introduced by
Derrida and Brunet \cite{Derrida05}; this model has also been
discussed by Visco \cite{Visco06} and Farago \cite{Farago02}.  The
model consists of a single Brownian particle or rod coupled to two
heat reservoirs at temperatures $T_1$ and $T_2$ with associated
damping constant $\Gamma_1$ and $\Gamma_2$. Here the heat $Q$ is
transported from one reservoir to the other via a single particle.
These authors find that the large deviation function has the
explicit form
\begin{eqnarray}
\mu(\lambda)= \frac{1}{2}\left[\Gamma_1 +\Gamma_2- \sqrt{
\Gamma_1^2+\Gamma_2^2+2\Gamma_1\Gamma_2(1-2\lambda T_1+2\lambda
T_2-2\lambda^2T_1T_2)}\right]. \label{ldf}
\end{eqnarray}
This expression for $\mu(\lambda)$ is consistent with the boundary
condition $\mu(0)=0$ following from (\ref{char}) and in accordance
with the fluctuation theorem (\ref{ft2}). i.e.,
$\mu(\lambda)=\mu(-\lambda+1/T_1-1/T_2)$. For $T_1=T_2$ the large
deviation function $\mu(\lambda)$ is symmetric, i.e.,
$\mu(\lambda)=\mu(-\lambda)$. In this case the heat fluctuates
between the two reservoirs and there is no net mean current. If we
decouple one of the reservoirs by setting $\Gamma_2=0$ (or
$\Gamma_1=0$) the system is in equilibrium with a single reservoir
and we have $\mu(\lambda)=0$ for all $\lambda$. Finally, from
(\ref{char}) we infer the mean value (the first cumulant) and the
second cumulant
\begin{eqnarray}
\frac{\langle Q\rangle}{t}&&=
(T_1-T_2)\frac{\Gamma_1\Gamma_2}{\Gamma_1+\Gamma_2}, \label{mean}
\\
\frac{\langle Q^2\rangle-\langle Q\rangle^2}{t}&&=
\frac{2\Gamma_1\Gamma_2T_1T_2}{\Gamma_1+\Gamma_2}+
\frac{2\Gamma_1^2\Gamma_2^2(T_1-T_2)^2}{(\Gamma_1+\Gamma_2)^3}.
\label{com}
\end{eqnarray}

Here we extend the Derrida-Brunet model to a Brownian particle
moving in a harmonic trap and analyze  the large deviation function.
The paper is organized in the following manner. In Sec.~\ref{model}
we set up the model with focus on the heat transfer $Q(t)$ and the
large deviation function $\mu(\lambda)$. In Sec.~\ref{analysis} we
evaluate the first and second cumulants within a Langevin approach,
comment of the Fokker-Planck approach but focus in particular on the
Derrida-Brunet method. We derive the differential equation for the
characteristic function and determine the large deviation function.
In Sec.~\ref{numsim} we support the analytical findings by a
numerical simulation. Sec.~\ref{disc} is devoted to a summary and a
discussion.
%%%%%%%%%%%%%%%%%%%%%%%%%%%%%%%%%%%%%%%%%%%%%%%%%%%%%%%%%
%%%%%%%%%%%%%%%%%%%%%%%%%%%%%%%%%%%%%%%%%%%%%%%%%%%%%%%%%
%%%%%%%%%%%%%%%%%%%%%%%%%%%%%%%%%%%%%%%%%%%%%%%%%%%%%%%%%
\section{\label{model} Model}
\begin{figure}
\includegraphics[width=0.5\hsize]{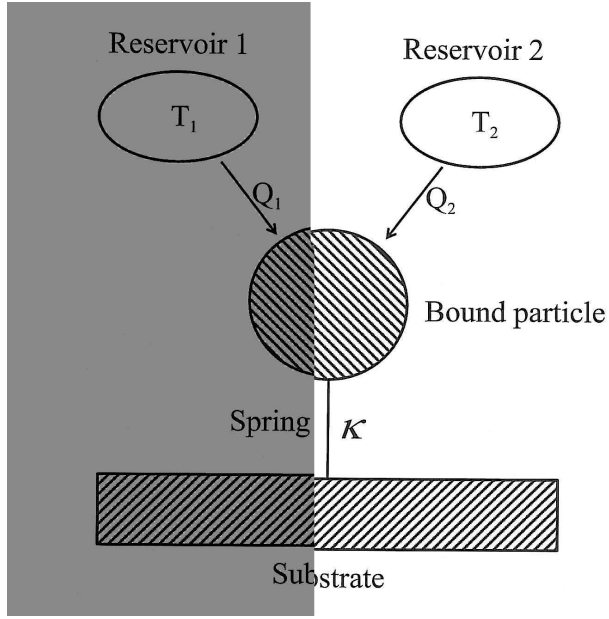}
\caption{We depict a harmonically bound particle interacting with
heat reservoirs at temperatures $T_1$ and $T_2$. The heat
transferred to the particle is denoted $Q_1$ and $Q_2$,
respectively. The particle is attached to a substrate with a
harmonic spring with force constant $\kappa$. } \label{fig1}
\end{figure}

%%%%%%%%%%%%%%%%%%%%%%%%%%%%%%%%%%%%%%%%%%%%%%%%%%%%%%%%%
%%%%%%%%%%%%%%%%%%%%%%%%%%%%%%%%%%%%%%%%%%%%%%%%%%%%%%%%%
%%%%%%%%%%%%%%%%%%%%%%%%%%%%%%%%%%%%%%%%%%%%%%%%%%%%%%%%%
We consider a 1D Brownian particle harmonically coupled to a
substrate by a force constant $\kappa$. This configuration also
corresponds to a Brownian particle in a harmonic trap. The particle
is, moreover, in thermal contact with two distinct heat reservoirs
at temperatures $T_1$ and $T_2$. The heat transferred in time $t$
from the two heat reservoirs is denoted $Q_1$ and $Q_2$,
respectively. Finally, the corresponding damping constants are
denoted $\Gamma_1$ and $\Gamma_2$, respectively. The configuration
is depicted in Fig.~\ref{fig1}.
Denoting the position of the particle by $u$ and the momentum by $p$
and assuming $m=1$, a conventional stochastic Langevin description
yields the equation of motion
\begin{eqnarray}
&&\frac{du}{dt}=p, \label{eq1}
\\
&&\frac{dp}{dt}=-(\Gamma_1+\Gamma_2)p-\kappa u+\xi_1+\xi_2,
\label{eq2}
\end{eqnarray}
where the Gaussian white noises $\xi_1$ and $\xi_2$ are correlated
according to
\begin{eqnarray}
&&\langle\xi_1(t)\xi_1(0)\rangle=2\Gamma_1T_1\delta(t), \label{n1}
\\
&&\langle\xi_2(t)\xi_2(0)\rangle=2\Gamma_2T_2\delta(t), \label{n2}
\\
&&\langle\xi_1(t)\xi_2(0)\rangle=0. \label{n3}
\end{eqnarray}
The heat flux from the reservoir at temperature $T_1$, i.e., the
rate of work done by the stochastic force $-\Gamma_1p+\xi_1$ on the
particle, is given by
\begin{eqnarray}
\frac{dQ_1}{dt}=-\Gamma_1p^2+p\xi_1; \label{flux1}
\end{eqnarray}
correspondingly, the heat flux from the reservoir at temperature
$T_2$ has the form
\begin{eqnarray}
\frac{dQ_2}{dt}=-\Gamma_2p^2+p\xi_2. \label{flux2}
\end{eqnarray}
The equations (\ref{eq1}-\ref{flux2}) define the problem and the
issue is to determine the asymptotic long time distribution for the
transferred heats $Q_1$ and $Q_2$,
\begin{eqnarray}
Q_n(t)=\int_0^td\tau(-\Gamma_np^2(\tau)+p(\tau)\xi_n(\tau)),~~n=1,2.
\label{heat}
\end{eqnarray}
At long times the heat distribution in terms of its characteristic
functions is given by (\ref{char}), i.e.,
\begin{eqnarray}
\langle e^{\lambda Q_n(t)}\rangle\propto
e^{t\mu_n(\lambda)},~~n=1,2, \label{char2}
\end{eqnarray}
where the large deviation function $\mu_n(\lambda)$ is associated
with $Q_n(t)$.

Noting that since the total noise $\xi=\xi_1+\xi_2$ is correlated
according to
$\langle\xi(t)\xi(0)\rangle=(2\Gamma_1T_1+2\Gamma_2T_2)\delta(t)$
and invoking the fluctuation-dissipation theorem \cite{Reichl98} we
readily infer that the system is in fact in equilibrium with the
effective temperature
$T=(\Gamma_1T_1+\Gamma_2T_2)/(\Gamma_1+\Gamma_2)$. This argument
also implies that the stationary distributions for $u$ and $p$ are
given by the Boltzmann-Gibbs expressions
$P_0(p)\propto\exp(-p^2/2T)$ and $P_0(u)\propto\exp(-\kappa
u^2/2T)$. The non-equilibrium features are obtained by splitting the
effective heat reservoir at temperature $T$ in two distinct heat
reservoirs at temperatures $T_1$ and $T_2$ and monitoring the heat
transfer.
%
%
%REV
From the equations of motion (\ref{eq1}) and (\ref{eq2}) we infer
two characteristic inverse lifetime in the system given by
$\Gamma_1+\Gamma_2$ and $\kappa^{1/2}$. In the following we assume
that the system is in a stationary non-equilibrium state at times
much larger than $(\Gamma_1+\Gamma_2)^{-1}$ and $\kappa^{-1/2}$ and
thus ignore initial conditions, i.e., the preparation of the system.
The role of the initial condition on the distribution $P(Q,t)$ is a
more technical issue, see Visco \cite{Visco06}.
%%%%%%%%%%%%%%%%%%%%%%%%%%%%%%%%%%%%%%%%%%%%%%%%%%%%%%%%%
%%%%%%%%%%%%%%%%%%%%%%%%%%%%%%%%%%%%%%%%%%%%%%%%%%%%%%%%%
%%%%%%%%%%%%%%%%%%%%%%%%%%%%%%%%%%%%%%%%%%%%%%%%%%%%%%%%%
\section{\label{analysis} Analysis}
%%%%%%%%%%%%%%%%%%%%%%%%%%%%%%%%%%%%%%%%%%%%%%%%%%%%%%%%%
%%%%%%%%%%%%%%%%%%%%%%%%%%%%%%%%%%%%%%%%%%%%%%%%%%%%%%%%%
%%%%%%%%%%%%%%%%%%%%%%%%%%%%%%%%%%%%%%%%%%%%%%%%%%%%%%%%%
We wish to address the issue to what extent the presence of the
spring represented by the term $\kappa u$ in the equation of
motion (\ref{eq2}) changes the large deviation function
(\ref{ldf}) in the free case.
%
%
%REV
In the case of an extended system coupled to heat reservoirs at
the edges, e.g., an harmonic chain, the heat is transported
deterministically across the system and the large deviation
function will depend on the internal structure of the system,
e.g., in the harmonic chain the spring constant. For vanishing
coupling the edges in contact with the reservoirs are disconnected
and the large deviation function must vanish. However, for a
single particle in a harmonic well there is no internal structure
or internal degrees of freedom and the case is special.

In addition to numerical simulations three analytical approaches are
available in investigating this issue: i) a Langevin equation method
taking its starting point in the equations of motion
(\ref{eq1}-\ref{eq2}) and determining the distribution of the
composite quantity $Q_n$ on the basis of a Greens function solution
and Wick's theorem, ii) an analysis based on the Fokker-Planck
equation for the joint distribution $P(u,p,Q,t)$, and iii) a direct
approach suggested by Derrida and Brunet which directly aims at
determining the long time behavior of the characteristic function
$\langle\exp(\lambda Q(t)\rangle$, yielding the large deviation
function.
%%%%%%%%%%%%%%%%%%%%%%%%%%%%%%%%%%%%%%%%%%%%%%%%%%%%%%%%%
%%%%%%%%%%%%%%%%%%%%%%%%%%%%%%%%%%%%%%%%%%%%%%%%%%%%%%%%%
%%%%%%%%%%%%%%%%%%%%%%%%%%%%%%%%%%%%%%%%%%%%%%%%%%%%%%%%%
\subsection{\label{langevin} Langevin approach}
%%%%%%%%%%%%%%%%%%%%%%%%%%%%%%%%%%%%%%%%%%%%%%%%%%%%%%%%%
%%%%%%%%%%%%%%%%%%%%%%%%%%%%%%%%%%%%%%%%%%%%%%%%%%%%%%%%%
%%%%%%%%%%%%%%%%%%%%%%%%%%%%%%%%%%%%%%%%%%%%%%%%%%%%%%%%%
Here we delve into the Langevin approach and discuss the
evaluation of the first two cumulants of the distribution
$P(Q,t)$.
%%%%%%%%%%%%%%%%%%%%%%%%%%%%%%%%%%%%%%%%%%%%%%%%%%%%%%%%%
%%%%%%%%%%%%%%%%%%%%%%%%%%%%%%%%%%%%%%%%%%%%%%%%%%%%%%%%%
%%%%%%%%%%%%%%%%%%%%%%%%%%%%%%%%%%%%%%%%%%%%%%%%%%%%%%%%%
\subsubsection{The first cumulant - The mean value}
%%%%%%%%%%%%%%%%%%%%%%%%%%%%%%%%%%%%%%%%%%%%%%%%%%%%%%%%%
%%%%%%%%%%%%%%%%%%%%%%%%%%%%%%%%%%%%%%%%%%%%%%%%%%%%%%%%%
%%%%%%%%%%%%%%%%%%%%%%%%%%%%%%%%%%%%%%%%%%%%%%%%%%%%%%%%%
The linear equations of motion (\ref{eq1}-\ref{eq2}) readily yield
to analysis. In Laplace space, defining $u(s)=\int_0^\infty dt
u(t)\exp(-st)$, etc., we obtain the solution
\begin{eqnarray}
p(s)=G(s)(\xi_1(s)+\xi_2(s)), \label{psol}
\end{eqnarray}
where the Greens function $G(s)$, broken up in normal mode
contributions, has the form
\begin{eqnarray}
G(s)=\frac{m_1}{s-s_1}+\frac{m_2}{s-s_2}. \label{green}
\end{eqnarray}
Here the resonances are given by
\begin{eqnarray}
&&s_1=-\frac{1}{2}[\Gamma+\tilde\Gamma], \label{res1}
\\
&&s_2=-\frac{1}{2}[\Gamma-\tilde\Gamma], \label{res2}
\\
&&\Gamma=\Gamma_1+\Gamma_2, \label{dam}
\\
&&\tilde\Gamma=\sqrt{\Gamma^2-4\kappa}; \label{dam2}
\end{eqnarray}
we note the relations $s_1+s_2=-\Gamma$, $s_1s_2=\kappa$, and
$s_1-s_2=-\sqrt{\Gamma^2+4\kappa}$.

The the amplitudes $m_1$ and $m_2$ have the form
\begin{eqnarray}
&&m_1=\frac{s_1}{s_1-s_2}, \label{amp1}
\\
&&m_2=\frac{s_2}{s_2-s_1}; \label{amp2}
\end{eqnarray}
note the sum rule $m_1+m_2$. For $\Gamma^2> 4 \kappa$ the system
is overdamped; for $\Gamma^2<4 \kappa$ the system exhibits a
damped oscillatory behavior with frequency
$\sqrt{4\kappa-\Gamma^2}$. In time we infer the solution
\begin{eqnarray}
p(t)=\int^t_0d\tau(m_1e^{s_1(t-\tau)}+
m_2e^{s_2(t-\tau)})(\xi_1(\tau)+\xi_2(\tau)). \label{mom}
\end{eqnarray}
We note that in the limit $\kappa\rightarrow 0$, $s_1\rightarrow
-\Gamma$, $s_2\rightarrow 0$, $m_1\rightarrow 1$, and
$m_2\rightarrow 0$, the position $u$ is decoupled from the momentum
$p$ and we recover the model proposed by Derrida and Brunet
\cite{Derrida05}.

Expressing time integration as a matrix multiplication and
introducing the short hand notation $p=(G_1+G_2)(\xi_1+\xi_2)$,
where $G_n(t, t')=m_n\exp(s_n(t-t'))\eta(t-t'),~n=1,2$, we obtain
from (\ref{flux1}-\ref{flux2})
\begin{eqnarray}
\frac{dQ_n}{dt}=-\Gamma_n((G_1+G_2)(\xi_1+\xi_2))^2+
\xi_n(G_1+G_2)(\xi_1+\xi_2). \label{flux3}
\end{eqnarray}
For the mean flux $d\langle Q_n\rangle/dt$ we then have averaging
over the noises $\xi_1$ and $\xi_2$ according to
(\ref{n1}-\ref{n3})
\begin{eqnarray}
\frac{d\langle Q_n\rangle}{dt}=
-\Gamma_n(2\Gamma_1T_1+2\Gamma_2T_2)(G_1+G_2)^2+
2\Gamma_nT_n(G_1(0)+G_2(0)). \label{flux4}
\end{eqnarray}
Inserting $\int G_n(t-t')^2dt'=-m_n^2/2s_n$, $\int
G_1(t-t')G_2(t-t')dt'=-m_1m_2/(s_1+s_2)$,  and
$G_n(0)=m_n\eta(0)$, $\eta(0)=1/2$, and reducing the expression we
obtain
\begin{eqnarray}
\frac{d\langle Q_n\rangle}{dt}=
2\Gamma_n(\Gamma_1T_1+\Gamma_2T_2)\left(\frac{m_1^2}{2s_1}+
\frac{m_2^2}{2s_2}+\frac{2m_1m_2}{s_1+s_2}\right)+\Gamma_nT_n(m_1+m_2).
\label{flux5}
\end{eqnarray}
By insertion of $m_1$, $m_2$,$s_1$, and $s_2$ the dependence on
the spring constant $\kappa$ cancels out and we obtain
\begin{eqnarray}
\frac{\langle
Q_1\rangle}{t}=\frac{\Gamma_1\Gamma_2}{\Gamma_1+\Gamma_2}(T_1-T_2),
\label{mean3}
\\
\frac{\langle
Q_2\rangle}{t}=\frac{\Gamma_1\Gamma_2}{\Gamma_1+\Gamma_2}(T_2-T_1),
\label{mean4}
\end{eqnarray}
independent of $\kappa$ and in agreement with the free particle
case (\ref{mean}). The independence of the mean value shows that
the heat transport is unaffected by the presence of the spring.
This feature is a result of the absence of internal structure in
the single particle case.
%%%%%%%%%%%%%%%%%%%%%%%%%%%%%%%%%%%%%%%%%%%%%%%%%%%%%%%%%
%%%%%%%%%%%%%%%%%%%%%%%%%%%%%%%%%%%%%%%%%%%%%%%%%%%%%%%%%
%%%%%%%%%%%%%%%%%%%%%%%%%%%%%%%%%%%%%%%%%%%%%%%%%%%%%%%%%
\subsubsection{ The second cumulant}
%%%%%%%%%%%%%%%%%%%%%%%%%%%%%%%%%%%%%%%%%%%%%%%%%%%%%%%%%
%%%%%%%%%%%%%%%%%%%%%%%%%%%%%%%%%%%%%%%%%%%%%%%%%%%%%%%%%
%%%%%%%%%%%%%%%%%%%%%%%%%%%%%%%%%%%%%%%%%%%%%%%%%%%%%%%%%
The evaluation of the second cumulant is more lengthy, involving
Wick's theorem \cite{Zinn-Justin89} applied to four noise
variables. Focussing on $Q=Q_1$ we have in matrix form
\begin{eqnarray}
\langle Q^2\rangle=\int^t_0 d\tau\int^t_0
d\tau'\langle(-\Gamma_1\xi GG\xi+\xi_1G\xi)(-\Gamma_1\xi
G'G'\xi+\xi_1G'\xi)\rangle, \label{Q2}
\end{eqnarray}
where $G=G_1+G_2$, $\xi=\xi_1+\xi_2$, $\xi GG\xi=\int
dt'dt''\xi(t')G(\tau,t')G(\tau,t'')\xi(t'')$, and $\xi G'G'\xi=\int
dt'dt''\xi(t')G(\tau',t')G(\tau',t'')\xi(t'')$. Applying Wick's
theorem to the product $\langle\xi\xi\xi\xi\rangle$ entering in
(\ref{Q2}) we note that only the pairwise contractions between the
$\tau$ and $\tau'$ factors in (\ref{Q2}) contribute to the cumulant
$\langle Q^2\rangle-\langle Q\rangle^2$; the contractions within the
$\tau$ and $\tau'$ terms factorize in (\ref{Q2}) and yield $\langle
Q\rangle^2$. Inserting $\xi=\xi_1+\xi_2$, applying Wick's theorem in
pairing the noise variables, and using (\ref{n1}-\ref{n3}), we
obtain
\begin{eqnarray}
&&\langle Q^2\rangle-\langle Q\rangle^2= \int^t_0
dt'\int^t_0dt''[L(t',t'')+M(t',t'')+N(t',t'')],
\\
&&L(t't'')=8\Gamma_1^2(\Gamma_1T_1+\Gamma_2T_2)^2 \int d\tau
d\tau' G(t',\tau)G(t',\tau')G(t'',\tau)G(t'',\tau'),
\\
&&M(t't'')= 4(\Gamma_1^2T_1^2+\Gamma_1\Gamma_2T_1T_2)
\delta(t'-t'')\int d\tau G(t',\tau)^2,
\\
&&N(t't'')=-8\Gamma_1(\Gamma_1^2T_1^2+\Gamma_1\Gamma_2T_1T_2)
G(t',t'')\int d\tau G(t',\tau)G(t'',\tau).
\end{eqnarray}
Finally, inserting $G=G_1+G_2$, using $G_n(t,
t')=m_n\exp(s_n(t-t'))\eta(t-t')$, and performing the integrations
over $t'$, $t''$, $\tau$, and $\tau'$, the dependence on the
spring constant again cancels out and we obtain the free particle
result
\begin{eqnarray}
\frac{\langle Q^2\rangle-\langle Q\rangle^2}{t}&&=
\frac{2\Gamma_1\Gamma_2T_1T_2}{\Gamma_1+\Gamma_2}+
\frac{2\Gamma_1^2\Gamma_2^2(T_1-T_2)^2}{(\Gamma_1+\Gamma_2)^3}.
\label{com1}
\end{eqnarray}
The Langevin approach turns out to be too cumbersome for the
present purposes and we shall not pursue it further but note that
the results for the two lowest cumulants corroborate the
suggestion that the large deviation function is independent of the
spring.
%%%%%%%%%%%%%%%%%%%%%%%%%%%%%%%%%%%%%%%%%%%%%%%%%%%%%%%%%
%%%%%%%%%%%%%%%%%%%%%%%%%%%%%%%%%%%%%%%%%%%%%%%%%%%%%%%%%
%%%%%%%%%%%%%%%%%%%%%%%%%%%%%%%%%%%%%%%%%%%%%%%%%%%%%%%%%
\subsection{\label{fokker-planck} Fokker-Planck approach}
%%%%%%%%%%%%%%%%%%%%%%%%%%%%%%%%%%%%%%%%%%%%%%%%%%%%%%%%%
%%%%%%%%%%%%%%%%%%%%%%%%%%%%%%%%%%%%%%%%%%%%%%%%%%%%%%%%%
%%%%%%%%%%%%%%%%%%%%%%%%%%%%%%%%%%%%%%%%%%%%%%%%%%%%%%%%%
Although we shall eventually complete the analysis using the
Derrida-Brunet method, we include for the benefit of the reader
and for completion the Fokker-Planck approach and the issues
arising in this context. It is here convenient to consider the
Fokker-Planck equation for the joint distribution $P(u,p,Q,t)$,
$Q=Q_1$. It has the form
\begin{eqnarray}
\frac{dP}{dt}=&&\{P,H\}+(\Gamma_1T_1+
\Gamma_1T_2)\frac{d^2P}{dp^2}+(\Gamma_1+\Gamma_2)\frac{d(pP)}{dp}
\nonumber
\\
&&+\Gamma_1\frac{d}{dQ}\left[(p^2+T_1)P+
T_1p^2\frac{dP}{dQ}+2T_1p\frac{dP}{dp}\right], \label{fp}
\end{eqnarray}
where $\{P,H\}$ denotes the Poisson bracket
\begin{eqnarray}
\{P,H\}=\frac{dP}{dp}\frac{dH}{du}-\frac{dP}{du}\frac{dH}{dp}=\kappa
u\frac{dP}{dp}-p\frac{dP}{du}. \label{pois}
\end{eqnarray}
The heat distribution after having analyzed the Fokker-Planck
equation is then given by
\begin{eqnarray}
P(Q,t)=\int dudp P(u,p,Q,t). \label{heat4}
\end{eqnarray}
Defining the characteristic function with respect to the  heat by
\begin{eqnarray}
C(\lambda)=\int dQ P(u,p,Q,t)e^{\lambda Q}, \label{char3}
\end{eqnarray}
and noting that $d/dQ\rightarrow -\lambda$ and
$d^2/dQ^2\rightarrow\lambda^2$ we obtain for $C$
\begin{eqnarray}
\frac{dC(\lambda)}{dt}=L(\lambda)C(\lambda), \label{fp2}
\end{eqnarray}
where the Liouville operator  $L$ has the form
\begin{eqnarray}
L(\lambda)C(\lambda)=&&\{C(\lambda),H\}+(\Gamma_1T_1+
\Gamma_1T_2)\frac{d^2C(\lambda)}{dp^2}+
(\Gamma_1+\Gamma_2)\frac{d(pC(\lambda))}{dp} \nonumber
\\
&&-\Gamma_1\lambda\left[(p^2+T_1)C(\lambda)- \lambda
T_1p^2C(\lambda)+2T_1p\frac{dC(\lambda)}{dp}\right].
\label{liouvl}
\end{eqnarray}

The case of an unbound particle Brownian particle for $\kappa=0$
has been discussed in detail by Visco \cite{Visco06}, see also
Farago \cite{Farago02}. Here $\{C(\lambda),H\}=-pdC(\lambda)/du$
and integrating over the position $u$ which is decoupled from the
momentum $p$ we obtain a second order differential equation for
$C$ of the Hermite type. By means of the transformation
\begin{eqnarray}
C(\lambda)=e^{-A(\lambda)p^2}\tilde C(\lambda),~~
A(\lambda)=\frac{\Gamma_1+
\Gamma_2-2\lambda\Gamma_1T_1}{4(\Gamma_1T_1+\Gamma_2T_2)},
\label{para}
\end{eqnarray}
$\tilde C(\lambda)$ satisfies the Schr\"odinger equation for a
harmonic oscillator and we infer the spectral representation
\begin{eqnarray}
C(\lambda)=e^{-A(\lambda)(p^2-p_0^2)}\sum_{n=0}
e^{E_n(\lambda)t}\Psi_n(p)\Psi_n(p_0), \label{spec}
\end{eqnarray}
where $-E_n(\lambda)$ is the discrete harmonic oscillator spectrum
and $\Psi_n(p)$ the associated normalized eigenfunctions. We have,
moreover, imposed the initial condition $C(t=0)=\delta(p-p_0)$,
where $p_0$ is the initial momentum. The large deviation function
is thus given by the ground state energy $-E_0(\lambda)$ yielding
(\ref{ldf}); for further discussion see Visco \cite{Visco06}.

In the case of a bound Brownian particle for $\kappa\ne0$ the
Poisson bracket enters and the position of the particle comes into
play. The Liouville operator becomes second order in $u$ and $p$
and is more difficult to analyze. We shall not pursue a further
analysis of the Fokker-Planck equation here but anticipate, in
view of the properties of the cumulants discussed above, that the
maximal eigenvalue yielding $\mu$ remains independent of $\kappa$.
%%%%%%%%%%%%%%%%%%%%%%%%%%%%%%%%%%%%%%%%%%%%%%%%%%%%%%%%%
%%%%%%%%%%%%%%%%%%%%%%%%%%%%%%%%%%%%%%%%%%%%%%%%%%%%%%%%%
%%%%%%%%%%%%%%%%%%%%%%%%%%%%%%%%%%%%%%%%%%%%%%%%%%%%%%%%%
\subsection{\label{derrida-brunet} Derrida-Brunet approach}
%%%%%%%%%%%%%%%%%%%%%%%%%%%%%%%%%%%%%%%%%%%%%%%%%%%%%%%%%
%%%%%%%%%%%%%%%%%%%%%%%%%%%%%%%%%%%%%%%%%%%%%%%%%%%%%%%%%
%%%%%%%%%%%%%%%%%%%%%%%%%%%%%%%%%%%%%%%%%%%%%%%%%%%%%%%%%
It is common to both the Langevin approach and the Fokker-Planck
approach that they carry a large overhead in the sense that one
addresses either the complete noise averaged solution of the
coupled equations of motion for $u$ and $p$ or the complete
distribution $P(u,p,Q,t)$. On the other hand, the method proposed
by Derrida and Brunet \cite{Derrida05} circumvent these issues and
directly addresses the large deviation function $\mu$.

Focussing again on $Q=Q_1$ the long time structure of the heat
characteristic function
\begin{eqnarray}
C(t)=\langle e^{\lambda Q(t)}\rangle\propto e^{t\mu(\lambda)},
\label{char4}
\end{eqnarray}
immediately implies that $C(t)$ satisfies the first order
differential equation
\begin{eqnarray}
\frac{dC(t)}{dt}=\mu(\lambda)C(t). \label{diffchar}
\end{eqnarray}
The task is thus reduced to constructing this equation and in the
process determine the large deviation function $\mu(\lambda)$.

In order to deal with the singular structure of the noise
correlations as expressed in (\ref{n1}-\ref{n3}) and avoid issues
related to stochastic differential equation \cite{Gardiner97}, it
is convenient to coarse grain time on a scale given by the
interval $\Delta t$ and introduce coarse grained noise variables
\begin{eqnarray}
&& F_1=\frac{1}{\Delta t}\int_t^{t+\Delta t}\xi_1(\tau)d\tau,
\label{cn1}
\\
&& F_2=\frac{1}{\Delta t}\int_t^{t+\Delta t}\xi_2(\tau)d\tau.
\label{cn2}
\end{eqnarray}
Since $\xi_1$ and $\xi_2$ are stationary random processes $F_1$
and $F_2$ are time independent. Moreover, we have $\langle
F_1\rangle=\langle F_2\rangle=\langle F_1 F_2\rangle=0$, and the
correlations
\begin{eqnarray}
&& \langle F_1^2\rangle=\frac{2\Gamma_1T_1}{\Delta t},
\label{nav1}
\\
&& \langle F_2^2\rangle=\frac{2\Gamma_2T_2}{\Delta t}.
\label{nav2}
\end{eqnarray}
The coarse graining in time allows us to construct a difference
equation for $C(t)$ for then at the end letting $\Delta
t\rightarrow 0$. Using the notation $p(t+\Delta t)=p'$, etc., we
thus obtain in coarse grained time from the equations of motion
(\ref{eq1}-\ref{eq2}) to $O(\Delta t)$
\begin{eqnarray}
&&u'=u+p\Delta t,\label{eqav1}
\\
&&p'=p+(-(\Gamma_1+\Gamma_2)p-\kappa u+F_1+F_2)\Delta
t.\label{eqav2}
\end{eqnarray}
For the heat increment we have from (\ref{heat})
\begin{eqnarray}
Q'=Q+\int_t^{t+\Delta t}d\tau (-\Gamma_1p(\tau)^2+p(\tau)F_1),
\label{heat2}
\end{eqnarray}
Since from (\ref{nav1}) $F_1$ is of order $(\Delta t)^{-1/2}$ we
must carry the expansion to $O((\Delta t^2))$ and we obtain
\begin{eqnarray}
Q'=Q+(F_1p-\Gamma_1p^2)\Delta t+\frac{1}{2}(F_1F_2+F_1^2)(\Delta
t)^2. \label{heat3}
\end{eqnarray}
We next proceed to derive a difference equation for $C$. This
procedure will in general produce correlations of the type
$\langle e^{\lambda Q}p^2\rangle$, $\langle e^{\lambda
Q}u^2\rangle$, and $\langle e^{\lambda Q}pu\rangle$ which are
effectively dealt with by considering the generalized
characteristic function
\begin{eqnarray}
C=\langle e^{K+\lambda Q}\rangle, \label{char5}
\end{eqnarray}
where $K$ is a bilinear form in $u$ and $p$
\begin{eqnarray}
K=\alpha p^2 + \beta u p+\gamma u^2. \label{kform}
\end{eqnarray}
This procedure is equivalent to considering the Fokker-Planck
equation for the joint distribution $P(u,p,Q,t)$ as discussed in the
previous subsection. The idea is to choose $K$, i.e., the parameters
$\alpha$, $\beta$, and $\gamma$, in such a way that the unwanted
correlations vanish yielding an equation for $C$. The conditions on
$K$ then yields the large deviation function $\mu$ directly.

Embarking on the actual procedure below, we introduce the notation
\begin{eqnarray}
&&K'=K+\Delta K, \label{k}
\\
&&Q'=Q+\Delta Q, \label{q}
\end{eqnarray}
where inserting (\ref{eqav1}) and (\ref{eqav2}) to order $\Delta
t$
\begin{eqnarray}
\Delta K= &&2\alpha p(-(\Gamma_1+\Gamma_2)p-\kappa u+F_1+F_2)\Delta
t \nonumber
\\
&&+\beta(p^2+u(-(\Gamma_1+\Gamma_2)p-\kappa u+F_1+F_2))\Delta t
\nonumber
\\
&&+2\gamma up\Delta t, \label{delk}
\\
\Delta Q= &&(F_1p-\Gamma_1p^2)\Delta
t+\frac{1}{2}(F_1F_2+F_1^2)(\Delta t)^2, \label{delq}
\end{eqnarray}
Inserting in $C'=\langle\exp(K'+\lambda Q')\rangle$ and expanding
to $O(\Delta t)$ we have
\begin{eqnarray}
C'=\langle e^{K+\lambda Q}[1+\Delta K+\lambda\Delta Q +
\frac{1}{2}(\Delta K+\lambda \Delta Q)^2]\rangle. \label{char6}
\end{eqnarray}
Using the identity $\langle F^2\exp(-F^2/2\Delta)\rangle=
\Delta\langle\exp(-F^2/2\Delta)\rangle$ we can average over $F_1$
and $F_2$ according to (\ref{nav1}) and (\ref{nav2}) inside the
noise average defining $C$. We obtain after some algebra
collecting terms to $O(\Delta t)$
\begin{eqnarray}
C'= C+\mu C\Delta t +\langle e^{K+\lambda Q}(Ap^2+Bpu+D
u^2)\rangle\Delta t, \label{char7}
\end{eqnarray}
where the intermediate parameters $A$, $B$, $D$ and $\mu$  in terms
of $\alpha$, $\beta$, $\gamma$ and $\lambda$ are given by
\begin{eqnarray}
&&A= 4\alpha^2(\Gamma_1T_1+\Gamma_2T_2)+
2\alpha(2\lambda\Gamma_1T_1-(\Gamma_1+\Gamma_2))+
\beta-\lambda\Gamma_1+\lambda^2\Gamma_1T_1, \label{a}
\\
&&B=-2\alpha\kappa-\beta(\Gamma_1+\Gamma_2-
2\lambda\Gamma_1T_1)+4\alpha\beta(\Gamma_1 T_1+\Gamma_2T_2)+2\gamma,
\label{b}
\\
&&D=\frac{1}{2}\beta^2-\beta\kappa, \label{c}
\\
&&\mu= 2\alpha(\Gamma_1T_1+\Gamma_2T_2)+ \lambda\Gamma_1T_1.
\label{d}
\end{eqnarray}
We note that the expression (\ref{char7}) involves correlations
between $\exp(K+\lambda Q)$ and $p^2$, $u^2$ and $pu$. However,
since $K$ is arbitrary we can obtain closure by choosing $K$,
i.e., $\alpha$, $\beta$ and $\gamma$, in such a manner that $A=0$,
$B=0$, and $D=0$. In the limit $\Delta t\rightarrow 0$
(\ref{char7}) then reduces to the differential equation
(\ref{diffchar}) and $\mu$ locks on to the large deviation
function

In the present case of a bound Brownian particle the discussion is
particularly simple. The condition $D=0$ immediately implies the two
solutions $\beta=0$ and $\beta=2\kappa$. However, since $\mu=0$ for
$\lambda=0$, the solution $\beta=2\kappa$ must be discarded and we
set $\beta=0$. Likewise, $\gamma$ is chosen so that $B=0$. Finally,
the condition $A=0$  yields a quadratic equation for $\alpha$ with
admissible solution
\begin{eqnarray}
\alpha(\lambda)= \frac{\Gamma_1+\Gamma_2-2\lambda\Gamma_1T_1-
\sqrt{(\Gamma_1+\Gamma_2)^2+2\Gamma_1\Gamma_2(1-2\lambda
T_1+2\lambda
T_2-2\lambda^2T_1T_2)}}{4(\Gamma_1T_1+\Gamma_2T_2)},\label{alsol}
\end{eqnarray}
and we recover the case (\ref{ldf}) for the free Brownian particle,
i.e.,
\begin{eqnarray}
\mu(\lambda)= \frac{1}{2}\left[\Gamma_1 +\Gamma_2- \sqrt{
\Gamma_1^2+\Gamma_2^2+2\Gamma_1\Gamma_2(1-2\lambda T_1+2\lambda
T_2-2\lambda^2T_1T_2)}~\right]. \label{ldf23}
\end{eqnarray}
%
%%%%%%%%%%%%%%%%%%%%%%%%%%%%%%%%%%%%%%%%%%%%%%%%%%%%%%%%%
%%%%%%%%%%%%%%%%%%%%%%%%%%%%%%%%%%%%%%%%%%%%%%%%%%%%%%%%%
%%%%%%%%%%%%%%%%%%%%%%%%%%%%%%%%%%%%%%%%%%%%%%%%%%%%%%%%%
\section{\label{numsim} Numerical simulations}
%%%%%%%%%%%%%%%%%%%%%%%%%%%%%%%%%%%%%%%%%%%%%%%%%%%%%%%%%
%%%%%%%%%%%%%%%%%%%%%%%%%%%%%%%%%%%%%%%%%%%%%%%%%%%%%%%%%
%%%%%%%%%%%%%%%%%%%%%%%%%%%%%%%%%%%%%%%%%%%%%%%%%%%%%%%%%
Here we perform a numerical simulation of eqs. (\ref{eq1})-(\ref{eq2}),
in order to sample the heat probability
distribution function (PDF) $P(Q,t)$ at long times and to verify that
the
distribution is independent of the spring constant $\kappa$ and in
conformity with the large deviation function $\mu$ given by
(\ref{ldf}).
Here and in the following the quantities will be expressed in
dimensionless units.

\begin{figure}
\includegraphics[width=0.5\hsize]{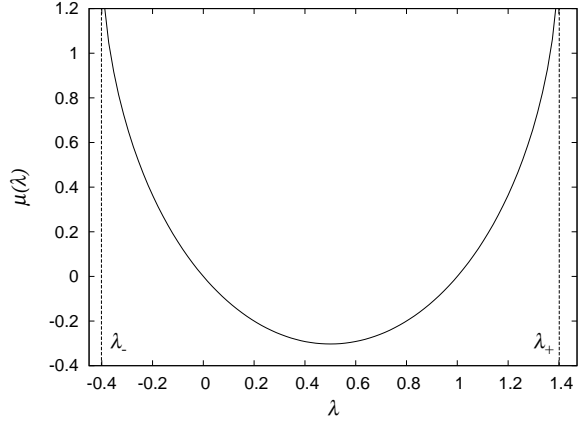}
\caption{Large deviation function $\mu(\lambda)$ as a function of
$\lambda$, as given by eq.~(\ref{ldf2}), for  $\Gamma_1=1$,
$\Gamma_2=2$, $T_1=1$, $T_2=2$. The shape is that of a half circle
lying between the branch points $\lambda_\pm$, as given by
(\ref{bp}).} \label{fig2}
\end{figure}

Following Visco \cite{Visco06}, see also
\cite{Derrida05,Lebowitz99}, $\mu(\lambda)$ can be expressed in
the form
\begin{eqnarray}
\mu(\lambda)=\frac{\Gamma_1+\Gamma_2}{2}-
\sqrt{\Gamma_1\Gamma_2T_1T_2}
\sqrt{(\lambda_+-\lambda)(\lambda-\lambda_-)}, \label{ldf2}
\end{eqnarray}
where the branch points are given by
\begin{eqnarray}
\lambda_\pm= \frac{1}{2}
\left[\frac{1}{T_1}-\frac{1}{T_2}\pm\sqrt{\left(\frac{1}{T_1}-
\frac{1}{T_2}\right)^2+\frac{(\Gamma_1+
\Gamma_2)^2}{\Gamma_1\Gamma_2T_1T_2}}~\right]; \label{bp}
\end{eqnarray}
note that $\lambda_+>0$ and $\lambda_-<0$. In Fig.~\ref{fig2} we
have depicted the large deviation function $\mu(\lambda)$ as a
function of $\lambda$.

\begin{figure}
\includegraphics[width=8cm]{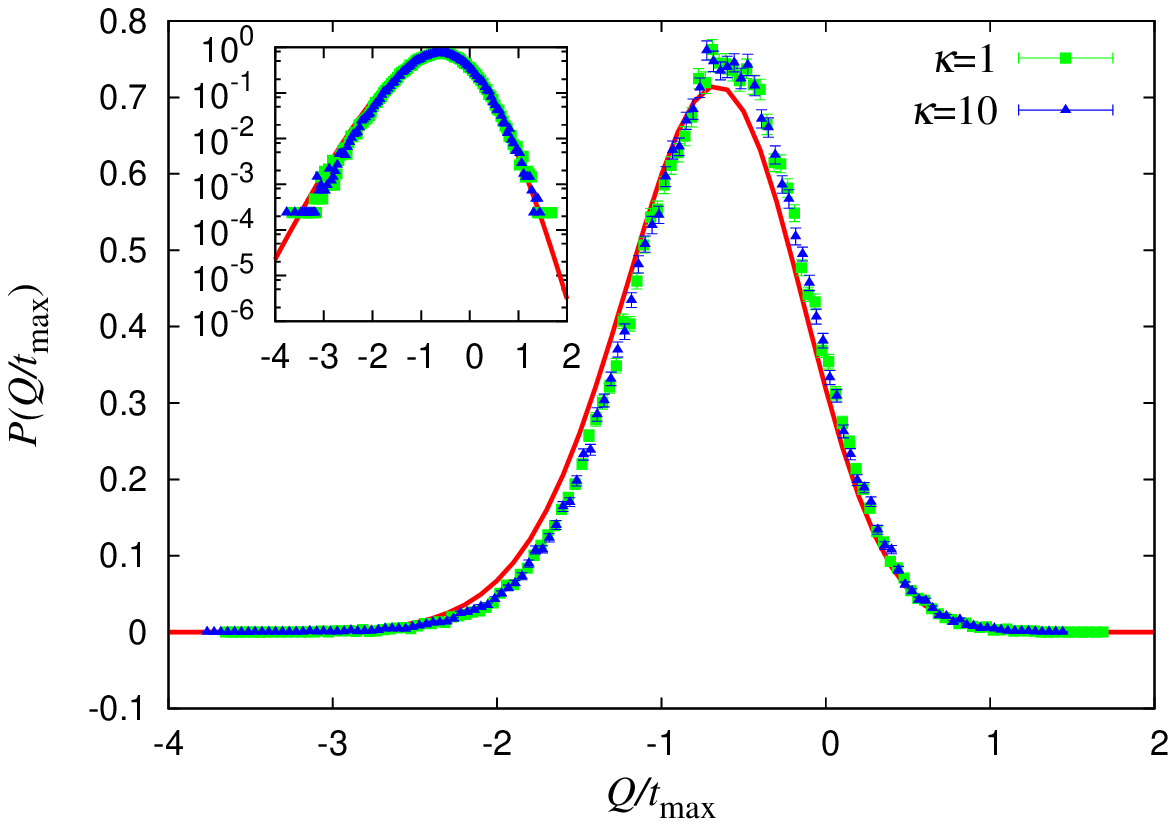}
\includegraphics[width=8cm]{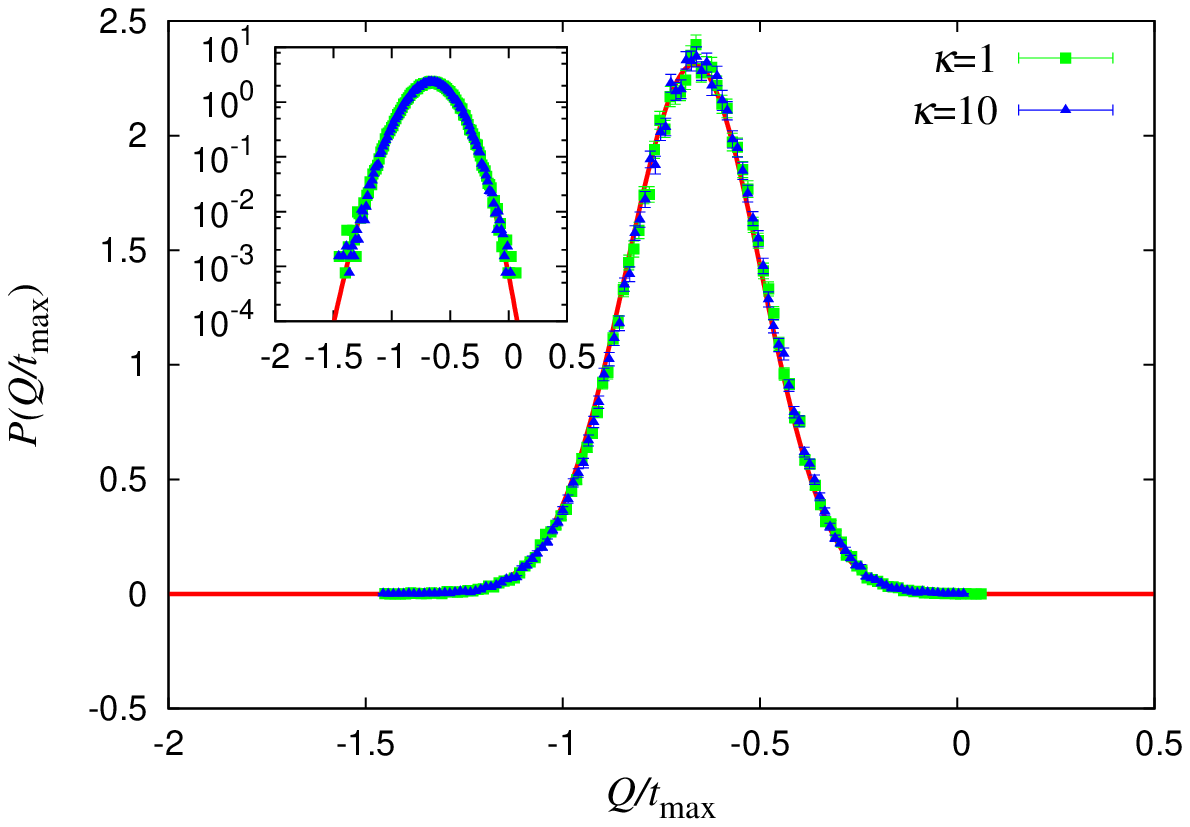}
\caption{Heat PDF $P(Q/t_{\text{max}})$ as a function of
$Q/t_{\text{max}}$ for $\Gamma_1=1$, $\Gamma_2=2$, $T_1=1$,
$T_2=2$ and two different values of $\kappa$ : left panel
$t_{\text{max}}=10$, right panel $t_{\text{max}}=100$. Full line:
theoretical prediction as given by (\ref{ldf22}). Linepoints: PDF
as obtained by  simulating $10^5$ independent trajectories. Inset:
log-linear plot.} \label{fig3}
\end{figure}

The large deviation function $F(q)$, $q=Q/t$, characterizing the
heat distribution, is determined parametrically from the large
deviation function $\mu(\lambda)$ according to the Legendre
transformation
\begin{eqnarray}
&&q=\mu'(\lambda) ~\rightarrow~\lambda^\ast=\lambda(q),
\label{par1}
\\
&&F(q)=\mu(\lambda^\ast)-\lambda^\ast\mu'(\lambda^\ast).
\label{par2}
\end{eqnarray}
We have, see also Visco \cite{Visco06},
\begin{eqnarray}
F(q)=\frac{1}{2}\left[\Gamma_1+\Gamma_2- q(\lambda_++\lambda_-)
-(\lambda_+-\lambda_-) \sqrt{\Gamma_1\Gamma_2T_1T_2+q^2}~\right],~~
\label{ldf33}
\end{eqnarray}
or inserting the branch points
\begin{eqnarray}
F(q)=\frac{1}{2}\left[\Gamma_1+\Gamma_2-
q\left(\frac{1}{T_1}-\frac{1}{T_2}\right)-\sqrt{\left(\frac{1}{T_1}-
\frac{1}{T_2}\right)^2+\frac{(\Gamma_1+
\Gamma_2)^2}{\Gamma_1\Gamma_2T_1T_2}}
\sqrt{\Gamma_1\Gamma_2T_1T_2+q^2}~\right].~~ \label{ldf22}
\end{eqnarray}
Inspection of this equation shows that for small $q$ we have a
displaced Gaussian distribution; for large $q$ we obtain exponential
tails originating from the branch points $\lambda_\pm$ in
$\mu(\lambda)$, i.e.,
\begin{eqnarray}
&&F(q)\sim -\lambda_+ q~~\text{for}~~q\gg 0,
\\
&&F(q)\sim -|\lambda_-| |q|~~\text{for}~~q\ll 0. \label{tails}
\end{eqnarray}
In Fig.~\ref{fig3} we have depicted the distribution
function $P(Q/t)\propto\exp(tF(Q/t))$, with $F(Q/t)$ given by
(\ref{ldf22}), as a function of $Q/t$ on linear scales and
log-linear scales (the inserts), for $\Gamma_1=1$, $\Gamma_2=2$,
$T_1=1$, $T_2=2$,  two different times
$t_{\text{max}}=10,\, 100$,  and two different values of
the force constant $\kappa=1,\, 10$. We find good agreement
between the simulations
and the analytical results for the ``central'' part of the
distribution. As expected, such an agreement improves as
$t_{\text{max}}$ increases, being excellent for $t_{\text{max}}=100$.
The tails cannot be sampled by the
simulations, as they correspond to rare trajectories, that would
require a very large simulation time to be observed.

\begin{figure}
%\psfrag{Q1}[cl][cl][0.75]{$\media{Q}$}
\psfrag{Q1}[cl][cl][0.75]{$m=1$}
%\psfrag{Q2}[cl][cl][0.75]{$\media{Q^2}$}
\psfrag{Q2}[cl][cl][0.75]{$m=2$}
%\psfrag{Q3}[cl][cl][0.75]{$\media{Q^3}$}
\psfrag{Q3}[cl][cl][0.75]{$m=3$}
%\psfrag{Q4}[cl][cl][0.75]{$\media{Q^4}$}
\psfrag{Q4}[cl][cl][0.75]{$m=4$}
\psfrag{e1}[cl][cl][0.75]{$\epsilon_1$}
\psfrag{e2}[cl][cl][0.75]{$\epsilon_2$}
\psfrag{e3}[cl][cl][0.75]{$\epsilon_3$}
\psfrag{e4}[cl][cl][0.75]{$\epsilon_4$}
\psfrag{ylab}[cc][cc][1.]{$\media{Q^m(\kappa)}/\media{Q^m(\kappa=0.01)}$}
\includegraphics[width=8cm]{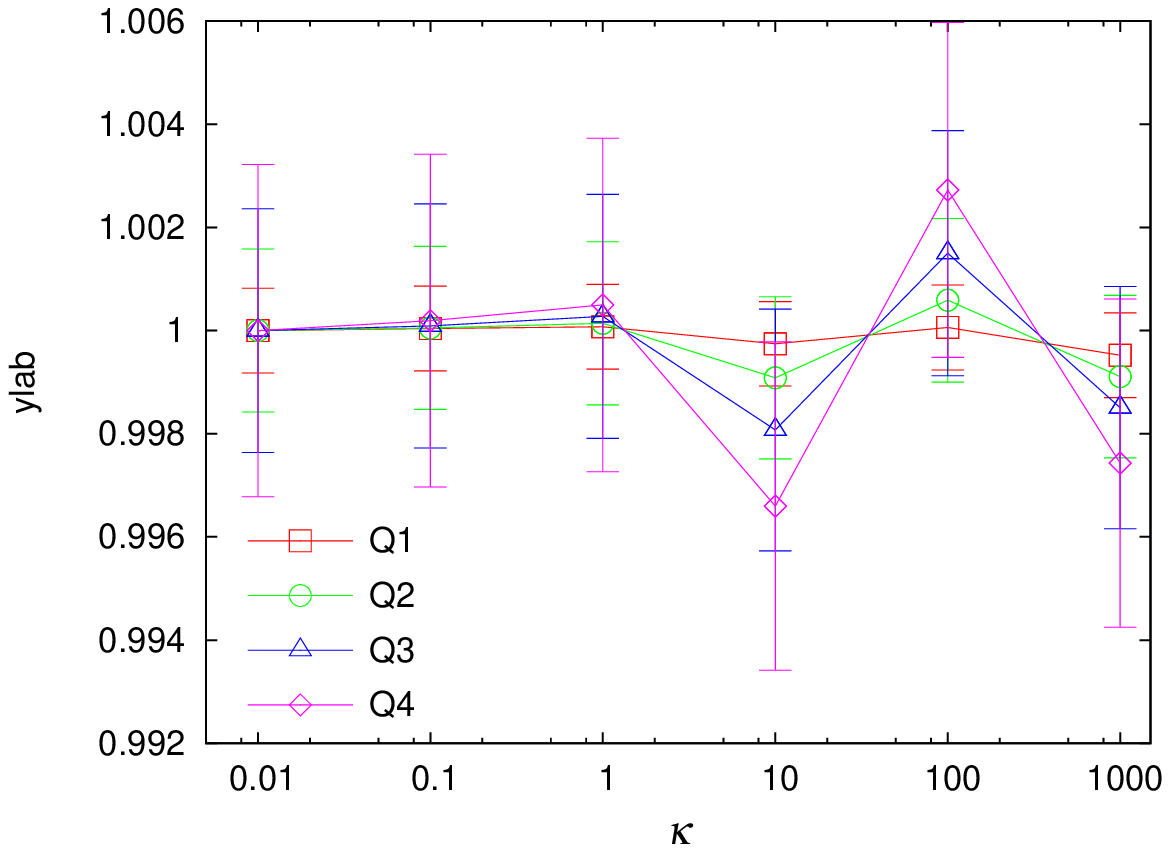}
\includegraphics[width=8cm]{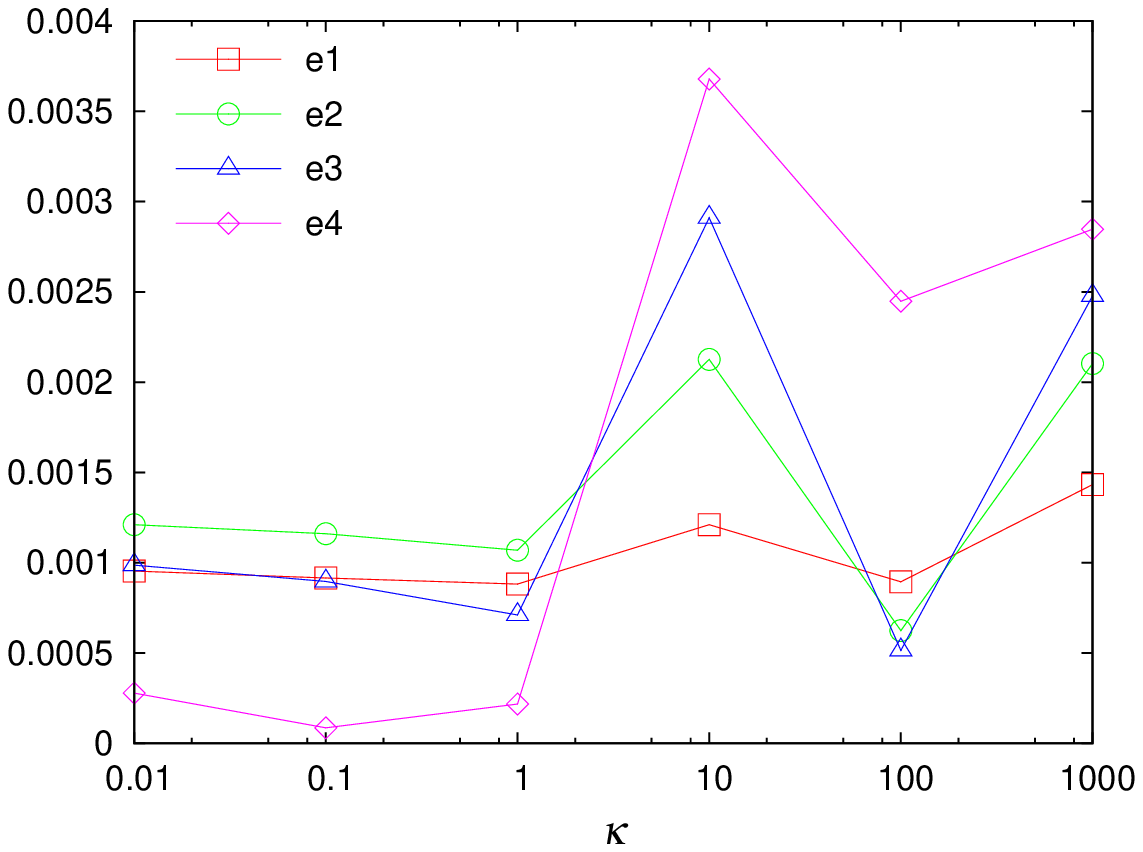}
\caption{Analysis of the first four moments as obtained by
numerical simulations with $t_{\text{max}}=100$, and  $10^5$
independent trajectories. Left panel relative change
$\media{Q^m(\kappa)}/\media{Q^m(\kappa=0.01)}$ of the first four
moments of the heat PDF as function of the spring constant
$\kappa$, wrt their value at $\kappa=0.01$. The moments are
practically constant over a range of six  orders of magnitude of
$\kappa$. Right panel: deviation of the first four moments from
the expected value $\epsilon_m$, as defined by (\ref{eqeps}).}
\label{fig5}
\end{figure}

To further support our main finding, namely that the heat PDF is
independent of the spring constant $\kappa$, we calculated the
first four moments of the distribution, over six orders on
magnitudes of $\kappa$, $10^{-2}\le \kappa\le10^4$. The
simulations were run for $t_{\text{max}}=100$, and  $10^5$
independent trajectories were sampled. The results are reported in
Fig.~\ref{fig5}. In the left panel we plot  the relative change
$\media{Q^m(\kappa)}/\media{Q^m(\kappa=0.01)}$, with $m=1\dots4$,
and we find that  the moments are practically constant over such a
large range of values of $\kappa$. Furthermore, for each value of
$\kappa$, we calculate the deviation $\epsilon_m$  of such moments
from the expected value  which reads:
\begin{equation}
\epsilon_m=\left|\frac{\media{Q_{\mathrm{num}}^m}-\media{Q_{\mathrm{ex}}^m}}{\media{Q^m_{\mathrm{ex}}}}\right|,
\label{eqeps}
\end{equation}
where $\media{Q_{\mathrm{num}}^m}$ is the $m$-th moment as
obtained by the numerical simulations, and
$\media{Q_{\mathrm{ex}}^m}$ is the corresponding exact value as
obtained by equation (\ref{ldf2}). The quantities  $\epsilon_m$
are plotted in the right panel of fig.~\ref{fig5}. We find, that
such deviations are negligible, basically due to numerical
imprecision.
%
%
%
%\begin{figure}
%\includegraphics[width=0.5\hsize]{tmax50.eps}
%\caption{Heat PDF $P(Q/t_{\text{max}})$ as a function of
%$Q/t_{\text{max}}$ for $\Gamma_1=1$, $\Gamma_2=2$, $T_1=1$,
%$T_2=2$,  $t_{\text{max}}=50$ and two different values of $\kappa$. Full line: theoretical prediction as given by eq.~(\ref{ldf22}). Linepoints: PDF as obtained by  simulating $10^4$ independent trajectories. Inset: log-linear plot.}
%\label{fig4}
%\end{figure}
%

\subsection{Numerical investigation of the fourth-order potential case}
In the present subsection, we investigate the heat PDF of a
particle coupled to the two heath baths at temperature $T_1$ and
$T_2$, but moving in a quadratic potential
\begin{equation}
V_4(u)=a_2 u^2+a_4 u^4.
\label{u4}
\end{equation}
Thus in (\ref{eq2}) the linear force is replaced by a term $2 a_2
u+4 a_4 u^3$. We sample the heat PDF by considering $10^5$
independent trajectories, with $ t_{\text{max}}=100$, and choose
different values for the parameters $a_2$ and $a_4$ in the
potential (\ref{u4}). The results for the first four moments are
reported in table~\ref{tab_u4}, and they provide a strong evidence
that also in this case the heat PDF, and so the large deviation
function, is independent of the details of the underlying
potential. As a bonus we also find that the first four moments are
well described by the same large deviation function that we
derived for the quadratic potential, which is independent of the
potential details indeed, in the present case of the parameter
$a_2$ and $a_4$ appearing in (\ref{u4}).
\begin{table}[h]
\caption{\label{tab_u4} Deviation $\epsilon_m$ of the first four
moments ((\ref{eqeps})) from the values predicted by the
substrate-independent large deviation function, (\ref{ldf2}). The
quantities $a_2$ and $a_4$ are the parameters of the fourth-order
potential $V_4$ as given by (\ref{u4}).}
\begin{tabular}{c|c|c|c|c|c}

%$a_2$ & $a_4$ & $\media{Q}$ & $\media{Q^2}$ & $\media{Q^3}$ &  $\media{Q^4}$ \\
$a_2$ & $a_4$ & $\epsilon_1$ & $\epsilon_2$ & $\epsilon_3$ &  $\epsilon_4$ \\
\hline
$-3$ & $1/2$ & $1.1 \times 10^{-3}$ & $1.9 \times 10^{-3}$ & $2.5 \times 10^{-3}$&  $3.0\times 10^{-3}$ \\
$-3/2$ & $1/12$ & $1.1 \times 10^{-3}$ & $1.6 \times 10^{-3}$ & $1.7\times 10^{-3}$&  $1.6 \times 10^{-3}$ \\
$1$ & $1$ & $1.0 \times 10^{-3}$ & $1.3 \times 10^{-3}$ & $1.3\times 10^{-3}$&  $1.0 \times 10^{-3}$ \\
\end{tabular}
\end{table}

%
%%%%%%%%%%%%%%%%%%%%%%%%%%%%%%%%%%%%%%%%%%%%%%%%%%%%%%%%%
%%%%%%%%%%%%%%%%%%%%%%%%%%%%%%%%%%%%%%%%%%%%%%%%%%%%%%%%%
%%%%%%%%%%%%%%%%%%%%%%%%%%%%%%%%%%%%%%%%%%%%%%%%%%%%%%%%%
\section{\label{disc} Discussion and conclusion}
%%%%%%%%%%%%%%%%%%%%%%%%%%%%%%%%%%%%%%%%%%%%%%%%%%%%%%%%%
%%%%%%%%%%%%%%%%%%%%%%%%%%%%%%%%%%%%%%%%%%%%%%%%%%%%%%%%%
%%%%%%%%%%%%%%%%%%%%%%%%%%%%%%%%%%%%%%%%%%%%%%%%%%%%%%%%%
In this paper we have discussed a bound Brownian particle coupled
to two distinct reservoirs, generalizing a model proposed by
Derrida and Brunet \cite{Derrida05}. The issue was to determine
whether the presence of a harmonic trap has an effect on the heat
transport between the reservoirs and on the large deviation
function characterizing the long time heat distribution function.
By a variety of analytical arguments based on a Langevin equation
evaluation of the two lowest cumulants and an evaluation of the
large deviation function by a direct method due to Derrida and
Brunet, supported by a numerical simulation, we have demonstrated
that the presence of a harmonic trap has no effect on the heat
distribution function which has the same form as in the unbound
case. This result is maybe intuitively evident since a single
particle, in contrast to an extensive system, does not have
internal degrees of freedom. Furthermore, we provide numerical
evidence, that the heat distribution function is unchanged if we
consider a fourth-order potential, again supporting our finding
that such a distribution is independent of the underlying
potential.

It also follows that the Gallavotti-Cohen fluctuation theorem
\cite{Gallavotti95} in (\ref{ft1}) is unchanged by the presence of
the spring. The fluctuation theorem is associated with the entropy
production $Q_1/T_1$ and $Q_2/T_2$ at the heat sources whereas the
presence of the spring represents a deterministic constraint not
associated with entropy production
\cite{Kurchan98,Derrida05,Lebowitz99}.

\acknowledgements We are grateful to C. Mejia-Monasterio for many
interesting discussions and for a critical reading of our
manuscript. We also thank A. Mossa, A. Svane, and U. Poulsen for
useful discussions. We thank B. Derrida and P. Visco for pointing
out to us ref.~\cite{Visco06}. We thank the Danish Centre for
Scientific Computing for providing us with computational
resources. The work of H. Fogedby has been supported by the Danish
Natural Science Research Council under grant no. 436246.

\bibliography{c:/user/manus/bib/articles,c:/user/manus/bib/books}
%\bibliography{toy}

%\bibliography{articles,books}
\end{document}